\documentclass[conference,letterpaper]{IEEEtran}

\usepackage[utf8]{inputenc} 
\usepackage[T1]{fontenc}
\usepackage{url}
\usepackage{ifthen}
\usepackage[cmex10]{amsmath} 

\usepackage[T1]{fontenc}
\usepackage[utf8]{inputenc}
\usepackage{bm}
\usepackage{amsmath}
\usepackage{amssymb,amsthm}
\usepackage{mathtools}
\usepackage{hyperref}
\usepackage[nameinlink]{cleveref}
\usepackage{color}
\usepackage{graphicx}
\graphicspath{{figures/}}
\newtheorem{theorem}{Theorem}
\newtheorem{definition}{Definition}
\newtheorem{corollary}{Corollary}

\newtheorem{proposition}{Proposition}
\usepackage{tikz}
\usetikzlibrary{backgrounds, calc, chains, fit, matrix, positioning, shadows, shapes, circuits.ee}
\tikzset{input/.style={}}
\tikzset{output/.style={}}
\tikzset{op/.style={circle, draw, thick, fill=black!10, minimum size=2.5ex, inner sep=0ex}}
\tikzset{filter/.style={rectangle, draw, thick, fill=black!10, minimum size=3.5ex, inner sep=1ex}}
\tikzset{nn/.style={trapezium, trapezium angle=80, draw, thick, fill=black!10, inner sep=1ex}}
\tikzset{branch/.style={circle, draw, thick, fill=black, minimum size=.5ex, inner sep=0ex}}
\tikzset{tensor/.style={rectangle, draw, thick, fill=white, minimum size=2em, double copy shadow={shadow xshift=.5ex,shadow yshift=-.5ex}}}
\tikzset{image/.style={rectangle, draw, thick, fill=white, minimum size=2em}}
\tikzset{>=direction ee}

\usepackage{pgfplots}
\pgfplotsset{compat=1.14}
\pgfplotsset{every axis/.append style={enlargelimits={abs=3pt},grid,axis lines=left}}
\pgfplotsset{every axis plot/.append style={thick,mark size=1.5pt,line join=bevel,mark options={solid}}}
\pgfplotsset{label style={font=\small}}
\pgfplotsset{tick label style={font=\footnotesize}}
\pgfplotsset{grid style={color=black!10}}
\pgfplotsset{legend style={draw=none,opacity=.85,font=\footnotesize,cells={anchor=west,opacity=1}}}
\pgfplotsset{every non boxed x axis/.style={xtick align=center,shorten <=-.5\pgflinewidth}}
\pgfplotsset{every non boxed y axis/.style={ytick align=center,shorten <=-.5\pgflinewidth}}
\pgfplotsset{every non boxed z axis/.style={ztick align=center,shorten <=-.5\pgflinewidth}}
\pgfplotsset{/pgf/number format/1000 sep={\,}}

\definecolor{gblue}{HTML}{1f77b4}
\definecolor{ggreen}{HTML}{2ca02c}

\DeclareMathOperator*{\cl}{cl}
\DeclarePairedDelimiterX{\divergence}[2]{[}{]}{#1\;\delimsize\|\;#2}

\newcommand{\markov}{\leftrightarrow}

\newcommand\comment[1]{}

\begin{document}
\title{The Rate-Distortion-Perception Tradeoff: \\ The Role of Common Randomness}

 \author{%
     \IEEEauthorblockN{Aaron B.~Wagner}
   \IEEEauthorblockA{School of Electrical and Computer Engineering \\
                     Cornell University \\
                     Ithaca, NY 14853 USA \\
                    Email: wagner@cornell.edu }
 }



\maketitle

\begin{abstract}
    A rate-distortion-perception (RDP) tradeoff has 
    recently been proposed by Blau and Michaeli and also
    Matsumoto. Focusing on the 
    case of perfect realism, which coincides with the problem
    of distribution-preserving lossy compression studied by
    Li \emph{et al.}, a coding theorem for the RDP tradeoff
    that allows for a specified amount of common randomness between the
    encoder and decoder is provided. The existing RDP tradeoff
    is recovered by allowing for
    the amount of common randomness to be infinite. The quadratic 
    Gaussian case is examined in detail.
\end{abstract}





%


\section{Introduction}

Classical rate-distortion theory seeks reconstructions
from rate-limited representations that are ``close'' to
the original source realization in a specified sense.
Closeness is conventionally measured via a single-letter
distortion measure, i.e., that one depends only on the
source and reconstruction realizations and that is 
additive over components of a 
string~\cite{Berger:RD,Shannon:IT,Shannon:RD}.

While broadly successful (e.g.~\cite{Pearlman:Said,Sayood:Compression}), 
this approach does have certain limitations. One is that the 
reconstruction might be qualitatively quite different from the source
realization that generated it. For an i.i.d.\ Gaussian 
source with mean-squared error (MSE) distortion measure,
the reconstruction is generally of lower 
power than the source. For stationary Gaussian sources,
the reverse waterfilling procedure~\cite{Berger:RD} generally
gives rise to reconstructions that have a null power
spectrum at high frequencies. Thus JPEG images look 
blurry at low bit-rates.

Of course, all distortion measures used in theoretical
studies of multimedia compression are proxies for the
measure of real interest, namely how the reconstruction
would be perceived by the (usually human) end-user.
In some cases, this end-user will prefer a reconstruction 
that is more distorted according to conventional 
measures. For instance, in some cases, MPEG Advanced
Audio Coding (ACC) populates high-frequency bands 
with artificial noise instead of leaving them 
null~\cite[Sec.~17.4.2]{Sayood:Compression}, in order to
match the power spectrum of the source;
this is termed \emph{Perceptual Noise
Substitution (PNS)}. This general idea has acquired renewed 
interest with the advent of neural-network-based image
compressors, 
for which powerful discriminators~\cite{Goodfellow:GANs,Arjovsky:WassersteinGAN,
Gulrajani:WassersteinGAN} 
can be used to encourage the compression system to output images 
that are indistinguishable from naturally-occurring 
ones~\cite{Tschannen:Preserving,Ripple:Realism,Agustsson:GAN}.

One way of capturing this notion mathematically is to
require that the distribution of reconstructions be
close, in some sense, to that of the source. Prior 
work has considered lossy compression under
such a constraint~\cite{Delp:Preserving,Li:Preserving:Long,
Li:Preserving:Long,Li:Preserving:Spectrum}.  
In particular, Li \emph{et al.}~\cite{Li:Preserving:Long}
consider the informational quantity
\begin{align}
    \nonumber
    R(\Delta) & = \inf_{P_{Y|X}} I(X;Y) \\
    \label{eq:BM}
    & \text{s.t.} \ E[D(X,Y)] \le \Delta \\
    \nonumber
    & \ \ \ \ \ Y \stackrel{d}{=} X,
\end{align}
which they call the \emph{distribution preserving
rate-distortion function (DP-RDF).} More recently,
Blau and Michaeli~\cite{Blau:Rate:Perception}
and Matsumoto~\cite{Matsumoto:RDP,Matsumoto:General}
consider a more general version that constrains
the divergence between the distributions of $X$
and $Y$ (see also~\cite{Blau:Tradeoff}), the former
calling it the \emph{rate-distortion-perception (RDP)} 
function. We shall adopt the latter nomenclature,
referring to the case in which $X \stackrel{d}{=} Y$
as the \emph{perfect realism} case. Both Li \emph{et al.}
and Blau and Michaeli provide a converse argument
in support of (\ref{eq:BM}). Li \emph{et al.}
provide an achievability argument in the Gaussian case.
L.~Theis and the author recently
provided an operational formulation and an achievability result
supporting the use of the RDP function~\cite{Theis:ICLR:2021}
(see also~\cite{Matsumoto:RDP}).
That formulation is variable-rate and assumes infinite common 
randomness between the encoder and the decoder.

This paper provides a coding theorem for a fixed-rate
scenario in which the amount of common randomness between
the encoder and decoder is constrained. It turns out that the
above RDP function only applies
when the available common randomness is infinite.
Thus, some care is required when interpreting
(\ref{eq:BM}) operationally.

The reason for characterizing the rate-distortion
tradeoff as a function of the amount of common
randomness available is not that common randomness
is a costly resource in compression scenarios \emph{per se}.
Indeed, in practice the encoder can include a small
seed for a pseudo-random number generator in its
message. It could even use the compressed representation
for one part of the source as the seed for another.
Rather, we note that randomization is not necessary
at all under conventional formulations of the problem:
in a fixed-rated setting\footnote{In some one-shot 
formulations, variable-rate codes can
benefit from common randomness as a form of time
sharing~\cite{Wagner:Sawbridge:DCC,Gyorgy:Uniform}.},
if the distortion is the average of a function that only 
depends on the realizations of the source
and the reconstruction,  then
in principle one could simply fix the realization
of the common randomness to be one that minimizes 
this average. Note that the distortion measure 
in question could be quite complex, such as
the median of the assessments of a 
collection of human subjects. Thus quantifying
the amount of randomness that is needed under
novel formulations is useful in that it
illustrates how much they depart from conventional
ones. The need for at least some
common randomness has already been illustrated
by Theis and Agustsson~\cite{Theis:Advantages}.
The precise characterization provided here
has the benefit of establishing an intimate 
connection between 
distribution-preserving compression and 
\emph{distributed channel synthesis} as studied by 
Cuff~\cite{Cuff:ChannelSim}. Indeed, the proof of our main result 
tracks that of~\cite[Theorem~II.1]{Cuff:ChannelSim}.

\section{Results for General Sources}

We are given a source distribution $P(x)$
over the alphabet $\mathcal{X}$ that
is assumed i.i.d.\ when extended to sequences, and
a distortion measure
\begin{equation}
    D: \mathcal{X} \times \mathcal{X} \mapsto [0,\infty).
\end{equation}
For a positive number $a$, let $[a]$ denote the set
$\{1,\ldots,\lfloor a \rfloor \}$.

\begin{definition}
    An $(n,2^{nR},2^{nR_c})$ \emph{code}
    consists of
    \begin{enumerate}
        \item[(a)] a (privately randomized) \emph{encoder}
            \begin{equation}
                F : \mathcal{X}^n \times [2^{n R_c}]
                   \mapsto [2^{nR}]
            \end{equation}
        \item[(b)] and a (privately randomized) \emph{decoder}
            \begin{equation}
                G : [2^{nR}] \times [2^{nR_c}]
                  \mapsto \mathcal{X}^n.
            \end{equation}
    \end{enumerate}
\end{definition}

\begin{definition}
   The triple
    $(R,R_c,\Delta)$ is \emph{achievable with near-perfect 
    (resp.\ perfect)
    realism} if for all $\epsilon > 0$, there exists a 
    sequence of codes, $\{(F_n,G_n)\}_{n = 1}^\infty$,
    the $n$th being
    $(n,2^{n(R+ \epsilon)},2^{n(R_c + \epsilon)})$,
    such that eventually we have 
    \begin{equation}
         E[D(X^n,Y^n)] \le \Delta + \epsilon
         \label{eq:Dconstraint}
    \end{equation}
     and
     \begin{align}
         d_{TV}(P_{X^n},P_{Y^n}) \le \epsilon, 
         \label{eq:dconstraint1}
         \intertext{(resp.}
         d_{TV}(P_{X^n},P_{Y^n}) = 0),
         \label{eq:dconstraint0}
     \end{align}
     where $Y^n = G_n(F_n(X^n,J),J)$ and $J$
     is uniformly distributed over $[2^{n(R_c + \epsilon)}]$,
     independent of the source. Here $d_{TV}(\cdot,\cdot)$
     refers to the total variation distance:
     \begin{equation}
         d_{TV}(P,Q) = \sup_{A} |P(A) - Q(A)|.
     \end{equation}
\end{definition}

Our first result shows that any sequence of codes that
achieves near-perfect realism can be upgraded to one
that achieves perfect realism with no asymptotic
change to the rates or the distortion. The result
holds under the following assumption on the 
distortion measure and source distribution pair.

\begin{definition}
    A distortion measure and source distribution pair
    $(D,P)$ is \emph{uniformly integrable} if 
    for every $\epsilon > 0$ there exists a $\delta > 0$
    such that
    \begin{equation}
        \sup_{X,Y,A} E[D(X,Y) \cdot 1_A] \le \epsilon,
        \label{eq:ui}
    \end{equation}
    where the supremum is over all random variables
    $X$ and $Y$ having marginal distribution $P$ and
    all events $A$ such that $\Pr(A) \le \delta$.
\end{definition}

If $\mathcal{X}$ is finite then evidently any pair of
distortion measure and source distribution is
uniformly integrable. The quadratic Gaussian case is
also uniformly integrable, since we have, by Cauchy-Schwarz,
\begin{multline}
    E[(X-Y)^2 1_A] \le (\sqrt{E[X^2 1_A]} +
                         \sqrt{E[Y^2 1_A]})^2,
\end{multline}
so since $X \stackrel{d}{=} Y$,
\begin{align}
    \sup_{X,Y,A} E[(X-Y)^2 1_A] &  \le
                         \sup_{A} 4 E[X^2 1_A] \\
                         & \le 4 \sqrt{E[X^4] \delta}\\
                         & = 4 \sqrt{3 \delta},
\end{align}
where we have used Cauchy-Schwarz for a second time.

\begin{theorem}
    \label{thm:equiv}
    If $(D,P)$ is uniformly integrable, then $(R,R_c,\Delta)$
    is achievable with perfect realism if and only if
    it is achievable with near-perfect realism.
\end{theorem}

Note that the proof does not rely on a 
single-letter characterization of the set of achievable 
rate-distortion triples.

\begin{IEEEproof}
    Suppose $(R,R_c,\Delta)$ is achievable with near-perfect
    realism. Fix $\epsilon > 0$ and choose $0 < \delta <
      \epsilon/2$ such that 
      \begin{equation}
          \label{eq:UIasapplied}
        \sup_{X,Y,A} E[D(X,Y) \cdot 1_A] \le \epsilon/2,
      \end{equation}
      where the supremum is over $X$ and $Y$ with
      marginals $P$ and events $A$ with probability
      at most $\delta$.
      Let $(F_n,G_n)$ be a
    sequence of codes, the $n$th being $(n,2^{n(R+\delta)},
    2^{n(R_c + \delta)})$ such that eventually
    \begin{equation}
        \label{eq:assumeddistortion}
        E[D(X^n,Y^n)] \le \Delta + \delta,
    \end{equation}
    and
    \begin{equation}
        \label{eq:assumeddivergence}
        d_{TV}(P_{X^n},P_{Y^n}) \le \delta,
    \end{equation}
    where $Y^n = G_n(F_n(X^n,J),J)$. For fixed message
    $i$ and common randomness $j$, the privately randomized
    decoder $G_n$ can be viewed as a conditional distribution
    \begin{equation}
        W(y^n|i,j) \quad y^n \in \mathcal{X}^n, 
             i \in [2^{n(R+\delta)}],
             j \in [2^{n(R_c+\delta)}].
    \end{equation}
    Let $PW(\cdot)$ denote
    the marginal distribution of the reconstruction induced
    by the encoder/decoder pair, i.e., for any
    event $A \subset \mathcal{X}^n$,
    \begin{equation}
        PW(A) = 
           \sum_{i,j} 
        \int_{\mathcal{X}^n}
           \frac{\Pr(F_n(x^n,j) = i)}
           {2^{n(R_c + \delta)}} 
               \; W(A|i,j) dP(x^n).
    \end{equation}
    By hypothesis, we have
    \begin{equation}
        d_{TV}(PW(y^n),P(y^n)) \le \delta.
    \end{equation}
    If the code does not already satisfy perfect realism,
    then we 
    leave the encoder untouched and replace the
    decoder $G_n$ with one, say $\tilde{G}_n$,
    with the same rates, nearly the same distortion,
    and perfect realism, as follows.

    Let $\Gamma(y^n)$ denote a probability distribution
    over $\mathcal{X}^n$ with respect to which 
    both $PW(y^n)$ and $P(y^n)$ are absolutely
    continuous (e.g., $(PW(y^n) + P(y^n))/2$). Define
    the set
    \begin{equation}
        \mathcal{X}^n_+ = \left\{y^n \in \mathcal{X}^n :
        \frac{dPW}{d\Gamma}(y^n) > \frac{dP}{d\Gamma}(y^n) \right\}.
    \end{equation}
    and the parameters
    \begin{equation}
        \theta_{y^n} =
            \frac{dP/d\Gamma(y^n)}
                {dPW/d\Gamma(y^n)} \quad y^n \in
                      \mathcal{X}_+^n.
    \end{equation}
    For any indices $i$ and $j$ and any
    set $A$ in $\mathcal{X}^n$ the
    alternate decoder $\tilde{G}_n$ is defined via
    the conditional distribution
    \begin{multline}
        \tilde{W}(A|i,j) = \\
        \int_{A \cap \mathcal{X}^n_+}
            \theta_{y^n} dW(y^n|i,j)
            + 
            W(A \backslash 
               \mathcal{X}^n_+|i,j)
                + \phi_{i,j} \cdot Q(A),
                \label{eq:altchannel}
    \end{multline}
    where the distribution $Q(y^n)$ is
    defined as
    \begin{equation}
        Q(A) = \frac{\int_A (dP/d\Gamma(y^n) - dPW/d\Gamma(y^n))^+ d\Gamma(y^n)}
        {\int (dP/d\Gamma(y^n) - dPW/d\Gamma(y^n))^+ d\Gamma(y^n)}.
    \end{equation}
    and the parameters $\phi_{i,j}$ are defined as
    \begin{equation}
        \phi_{i,j} =  \int_{\mathcal{X}^n_+}
            (1-\theta_{y^n}) dW(y^n|i,j).
    \end{equation}
    One can verify by direct calculation
    that $\tilde{W}(\cdot|i,j)$ is a
    probability distribution for each $i$ and $j$
    and moreover
    \begin{align}
        P\tilde{W}(A) & = 
           \sum_{i,j}
        \int_{\mathcal{X}^n}
           \frac{\Pr(F_n(x^n,j) = i)}
           {\lfloor 2^{n(R_c + \delta)}\rfloor}
               \; \tilde{W}(A|i,j) dP(x^n) \\
           & = P(A)
    \end{align}
    as desired. Thus $(F_n,\tilde{G}_n)$ achieves
    perfect realism. Let $Y^n$ denote the output
    of $G_n$ and $\tilde{Y}^n$ the output
    of $\tilde{G}_n$. By~(\ref{eq:altchannel}), we have
    that for each $i,j$, the total variation
    distance 
    satisfies
    \begin{equation}
        d_{TV}(W(\cdot|i,j),\tilde{W}(\cdot|i,j))
           \le \phi_{i,j}.
    \end{equation}
    Thus it is possible
    to couple $X^n$, $Y^n$, and $\tilde{Y}^n$ so
    that 
    \begin{equation}
        \Pr(Y \ne \tilde{Y}|J = j, F_n(X^n,J) = i)
        \le \phi_{i,j}.
    \end{equation}
    This in turn implies that
    \begin{align}
    & \Pr((X^n,Y^n) \ne (X^n,\tilde{Y}^n)) \\
        & = \sum_{i,j}
        \int_{\mathcal{X}^n}
          \frac{\Pr(F_n(x^n,j) = i)}
        {\lfloor 2^{n(R_c + \delta)} \rfloor}
          \phi_{i,j} dP(x^n) \\
        & = \sum_{i,j}
        \int_{\mathcal{X}^n} \int_{\mathcal{X}^n_+}
           \frac{\Pr(F_n(x^n,j) = i)}
        { \lfloor 2^{n(R_c + \delta)} \rfloor}
        \left(1 - \frac{\frac{dP}{d\Gamma}(y^n)}
                {\frac{dPW}{d\Gamma}(y^n)} \right) \\
                & \phantom{= \sum \int}
            dW(y^n|i,j) dP(x^n) \\
        & = \sum_{i,j}
        \int_{\mathcal{X}^n} \int_{\mathcal{X}^n_+}
           \frac{\Pr(F_n(x^n,j) = i)}
        {\lfloor 2^{n(R_c + \delta)} \rfloor} \\
        & \phantom{\sum\int } \frac{\frac{dPW}{d\Gamma}(y^n) - 
                \frac{dP}{d\Gamma}(y^n)}
                {\frac{dPW}{d\Gamma}(y^n)}
            dW(y^n|i,j) dP(x^n) \\
        & = \int_{\mathcal{X}^n_+}
        \frac{\frac{dPW}{d\Gamma}(y^n) - \frac{dP}{d\Gamma}(y^n)}
                {\frac{dPW}{d\Gamma}(y^n)}
            dPW(y^n) \\
        & = \int_{\mathcal{X}^n_+} \left[
            \frac{dPW}{d\Gamma}(y^n) - \frac{dP}{d\Gamma}(y^n) \right]
            d\Gamma(y^n) \\
        & = \int_{\mathcal{X}^n}
        \left[\frac{dPW}{d\Gamma}(y^n) - \frac{dP}{d\Gamma}(y^n)\right]^+
            d\Gamma(y^n) \\
        & = d_{TV}(P(\cdot),PW(\cdot)) \\
        & \le \delta.
    \end{align}
    This fact can then be used to bound the distortion
    achieved by $(F_n,\tilde{G}_n)$
    \begin{align}
        E[D(X^n,\tilde{Y}^n)] & = 
        \frac{1}{n} \sum_{i = 1}^n
             E[D(X_i,\tilde{Y}_i)] \\
             & = \frac{1}{n} \sum_{i = 1}^n
             E[D(X_i,\tilde{Y}_i) 1_{\tilde{Y}_i = Y_i}] \\
             & \phantom{= E[} + \frac{1}{n} \sum_{i = 1}^n
             E[D(X_i,\tilde{Y}_i) 1_{\tilde{Y}_i \ne Y_i}] \\
             & \le \Delta + \delta
             + \sup_{X,Y,A}  E[D(X,Y)1_A] \\
             & \le \Delta + \epsilon,
    \end{align}
   where the supremum is over all random variables
   $X$ and $Y$ having marginal distribution $P$ and
   all events $A$ such that $\Pr(A) \le \delta$ and
   we have used (\ref{eq:UIasapplied}), 
   (\ref{eq:assumeddistortion}),  and
   (\ref{eq:assumeddivergence}).
\end{IEEEproof}

Our main result is a characterization of the rate-distortion
tradeoff with perfect (or near-perfect) realism and limited 
common randomness.

\begin{definition}
    \begin{align}
        \mathcal{RD} = \Big\{(R,R_c,\Delta) & : \ \text{there exists}
        \ (U,Y) \ \text{such that} \\
         Y & \stackrel{d}{=} X \\
          X & \markov U \markov Y \\
         R & \ge I(X;U) \\ 
         R_c + R & \ge I(Y;U)  \\
         \Delta & \ge E[D(X,Y)] \Big\} 
    \end{align}
\end{definition}

\begin{theorem}
    If $(D,P)$ is uniformly integrable, then
    the triple
    $(R,R_c,\Delta)$ is achievable with perfect
    realism (or near-perfect realism) iff it is 
    contained in the closure of $\mathcal{RD}$.
    \label{thm:main}
\end{theorem}

    \begin{IEEEproof}
        Note that the equivalence between perfect and
        near-perfect realism follows from the previous
        result.
        Suppose $(R,R_c,\Delta)$ is achievable with
        perfect realism.
      Fix $\epsilon > 0$, and let 
      $(F_n, G_n)$ be a sequence of codes, the $n$th being
      $(n, 2^{n(R+\epsilon)}, 2^{n(R_c + \epsilon)})$ eventually
        satisfying
      (\ref{eq:Dconstraint}) and (\ref{eq:dconstraint0}). Fix $n$ 
       and let $I$ denote the message, i.e., 
      \begin{align}
          I & = F_n(X^n,J) \\
          Y^n & = G_n(F_n(X^n,J),J).
      \end{align}
        Let $T$ be uniformly distributed over $[n]$
      and $U = (T,I,J)$. Then we have
      \begin{align}
          n(R + \epsilon) & \ge H(I) \\
          & \ge H(I|J) \\
          & \ge I(X^n;I|J) \\
          & = I(X^n;I,J) \\
          & = \sum_{i = 1}^n I(X_i;I,J,X^{i-1}) \\
          & \ge \sum_{i = 1}^n I(X_i;I,J) \\
          & = n I(X_T;I,J|T) \\
          & = n I(X_T;U).
      \end{align}
        Similarly, since $Y^n \stackrel{d}{=} X^n$,
      \begin{align}
          n(R + R_c + 2\epsilon) & \ge H(I,J) \\
          & \ge I(Y^n;I,J) \\
          & \ge \sum_{i = 1}^n I(Y_i;I,J) \\
          & \ge n I(Y_T;I,J|T) \\
          & \ge n I(Y_T;U). 
      \end{align}
      Evidently we have $X_T \markov U \markov Y_T$,
      $X_T \stackrel{d}{=} Y_T \stackrel{d}{=} X$. 
      It is straightforward to verify that
      \begin{align}
          E[D(X_T,Y_T)] \le \Delta + \epsilon.
      \end{align}
      It follows that if $(R,R_c,\Delta)$ is achievable
        with perfect realism
        then it is within $\epsilon$ of $\mathcal{RD}$
        for any $\epsilon > 0$.

     For achievability, it suffices to show that
        $(R,R_c,\Delta)$ in the closure of $\mathcal{RD}$
        is achievable with near-perfect
     realism, by Theorem~\ref{thm:equiv}. Fix $\epsilon > 0$
     and $(U,Y)$ satisfying
     \begin{align}
         \label{eq:achieve0}
         R + \epsilon & > I(X;U) \\
         \label{eq:achieve1}
         R_c + R + 2\epsilon & > I(Y;U) \\
         \label{eq:achieve2}
         \Delta + \epsilon & > E[D(X,Y)]
     \end{align}
     and $X \markov U \markov Y$ and $Y \stackrel{d}{=} X$.
     Let $P(x,u,y)$ denote the joint distribution of these
     variables. Consider selecting a random codebook
     $U^n(i,j)$ for $i \in [2^{n(R+\epsilon)}]$,
     $j \in [2^{n(R_c+\epsilon)}]$ i.i.d. $P(u)$.
     By the \emph{soft covering lemma}~\cite[Lemma~IV.1]{Cuff:ChannelSim}, 
     (see also~\cite{Wyner:Common,Han:Output,Hayashi:General:Wiretap}) 
     and (\ref{eq:achieve1}) we have
     \begin{equation}
     \lim_{n \rightarrow \infty}
         E\left[d_{TV}\left(P_{Y^n},
            \sum_{i,j} \frac{P_{Y^n|U^n}(\cdot|U^n(i,j))}
         {\lfloor 2^{n(R + \epsilon)} \rfloor 
         \lfloor 2^{n(R_c + \epsilon)} \rfloor }\right)\right] = 0,
     \end{equation}
     and by (\ref{eq:achieve0})
     \begin{align}
         \nonumber
      &\lim_{n \rightarrow \infty}
         \frac{1}{2^{n(R_c + \epsilon)}}
           \sum_j E\left[d_{TV}\left[P_{X^n},
            \sum_{i} \frac{P_{X^n|U^n}(\cdot|U^n(i,j))}
           {\lfloor 2^{n(R + \epsilon)}\rfloor}\right] \right] \\
             & \phantom{\lim 2^{n(R_c + \epsilon}} = 0.
     \end{align}
     At the same time, by the law of large numbers, we have
     \begin{multline}
         \sum_{i,j}
             \int \int
             \frac{D(x^n,y^n)
             dP_{X^n,Y^n|U^n}(x^n,y^n|U^n(i,j)) }
               {\lfloor 2^{n(R+\epsilon)} \rfloor 
              \lfloor  2^{n(R_c+\epsilon)} \rfloor }
             \\
             \stackrel{P}{\rightarrow} E[D(X,Y)].
     \end{multline}
     Thus for all sufficiently large $n$, there exists a realization of the 
     code, $\{u^n(i,j)\}_{i,j}$, satisfying
     \begin{align}
         d_{TV}\left(P_{Y^n}(\cdot), \sum_{i,j}
         \frac{ P_{Y^n|U^n}(\cdot|u^n(i,j))}
         {\lfloor 2^{n(R +\epsilon)} \rfloor
         \lfloor 2^{n(R_c + \epsilon)} \rfloor }
         \right) & < \epsilon, \\
           \frac{1}{\lfloor 2^{n(R_c + \epsilon)} \rfloor}
           \sum_j d_{TV}\left(P_{X^n}(\cdot),
            \sum_{i} \frac{P_{X^n|U^n}(\cdot|u^n(i,j))}{\lfloor 2^{n(R + \epsilon)}
               \rfloor} 
             \right) & < \epsilon,
     \end{align}
     and
     \begin{multline}
         \sum_{i,j} \int \int
             \frac{D(x^n,y^n)
             dP_{X^n,Y^n|U^n}(x^n,y^n|u^n(i,j)) }
             {\lfloor 2^{n(R+\epsilon)} \rfloor
             \lfloor 2^{n(R_c+\epsilon)} \rfloor} 
             \\
                \le \Delta + \epsilon.
     \end{multline}
     Let $Q_{X^n,Y^n,I,J}$ denote the joint distribution
    \begin{multline}
        Q_{X^n,Y^n,I,J}(x^n,y^n,i,j) = \\
           \frac{1}{\lfloor 2^{n(R + \epsilon)} \rfloor
           \lfloor 2^{n(R_c + \epsilon)} \rfloor}
            P_{X^n,Y^n|U^n}(x^n,y^n|u^n(i,j)),
    \end{multline}
    and note that we have, eventually,
     \begin{align}
         \label{eq:TV1}
              d_{TV}(P_{Y^n},Q_{Y^n}) & < \epsilon \\
         \label{eq:TV2}
              \frac{1}{\lfloor 2^{n (R_c + \epsilon)} \rfloor}
         \sum_{j = 1}^{\lfloor 2^{n(R_c+\epsilon)}\rfloor} 
         d_{TV}(P_{X^n},Q_{X^n,J}(\cdot|j)) & < \epsilon, \\
         \label{eq:TV3}
                   E_Q[D(X^n,Y^n)] & < \Delta + \epsilon.
     \end{align}
     Given a source string $x^n$ and a realization $j$ of the common
     randomness, the encoder selects a message 
     $i \in [2^{n(R + \epsilon)}]$
     randomly, using private randomness, with probability
     \begin{equation}
         Q_{I|X^n,J}(i|x^n,j),
     \end{equation}
     assuming $Q(x^n,j) > 0$. Otherwise, it selects a message
     at random.
     The decoder creates $Y^n$ by passing $u^n(i,j)$ through
     the i.i.d.\ channel $P_{Y^n|U^n}$. 
     The resulting joint distribution is given by
     \begin{equation}
         dP_X(x^n) \frac{1}{\lfloor 2^{n(R_c + \epsilon)}\rfloor} 
            Q_{I|X^n,J}(i|x^n,j) dP_{Y^n|U^n}(y^n|u^n(i,j)),
     \end{equation}
     which we denote by $\tilde{Q}(\cdot)$.
     Now from (\ref{eq:TV1})-(\ref{eq:TV2}) we have
     that eventually
     (cf.\ Cuff~\cite[Eqs. (61) and (64)]{Cuff:ChannelSim})
     \begin{equation}
         \label{eq:TVclose}
         d_{TV}(Q,\tilde{Q}) < \epsilon.
     \end{equation}
     It follows by~(\ref{eq:TV1}) and the 
     triangle inequality for total variation distance
     that the code achieves near-perfect
     realism. At the same time, by (\ref{eq:TVclose}),
     \begin{align}
         E_{\tilde{Q}}[D(X^n,Y^n)] - E_{Q}[D(X^n,Y^n)] \le 
         \sup_{X,Y,A} E[D(X,Y) 1_A]
     \end{align}
     where the supremum is over $X$ and $Y$ 
     with marginal $P$ and event $A$ with
     probability at most $\epsilon$. The 
     conclusion then follows from (\ref{eq:TV3})
      and the uniform integrability assumption.
\end{IEEEproof}

The fact that we need only consider perfect realism
allows us to sidestep the continuity argument 
in the proof of~\cite[Thm.~II.1]{Cuff:ChannelSim}, which in 
turn allows us to establish the converse for general spaces.

The two extreme cases of Theorem~\ref{thm:main} are notable.
Formally substituting $R_c = \infty$ into $\mathcal{RD}$
yields the region in (\ref{eq:BM}). With no common
randomness, the region is different.

\begin{corollary}[No Common Randomness]
    The triple $(R,0,\Delta)$ is achievable iff
    $(R,\Delta)$ is contained in the closure of the set
    \begin{align}
        \mathcal{RD}_0 = \Big\{(R,\Delta) & : \ \text{there exists}
        \ (U,Y) \ \text{such that} \\
        Y & \stackrel{d}{=} X \\
        X & \markov U \markov Y \\
        R & \ge \max(I(X;U),I(U;Y)) \\
        \Delta & \ge E[D(X,Y)]\Big\}.
    \end{align}
\end{corollary}

\begin{IEEEproof}
    Evidently we have
    \begin{equation}
        \mathcal{RD}_0 = \{(R,\Delta) : (R,0,\Delta) \in \mathcal{RD}\},
    \end{equation}
    and thus
    \begin{align}
        \cl(\mathcal{RD}_0) & = \cl\{(R,\Delta) : (R,0,\Delta) \in \mathcal{RD}\} \\
          & \subseteq \{(R,\Delta) : (R,0,\Delta) \in \cl(\mathcal{RD}) \},
    \end{align}
    where $\cl(\cdot)$ denotes closure.
    Achievability then follows from Theorem~\ref{thm:main}. Conversely,  if $(R,0,\Delta)$
    is achievable, then for all $\epsilon > 0$, $(R+\epsilon,\epsilon,\Delta+\epsilon)
    \in \mathcal{RD}$. This implies that $(R+2\epsilon,0,\Delta+\epsilon) \in \mathcal{RD}$
    and hence $(R+2\epsilon,\Delta+\epsilon) \in \mathcal{RD}_0$,
    since the encoder can simply transmit some of its private randomness to the decoder.
    It follows that $(R,\Delta) \in \cl(\mathcal{RD}_0)$.
\end{IEEEproof} 

The quadratic Gaussian case, to which we turn next, 
illustrates the difference between these two cases.

\section{The Quadratic Gaussian Case}

\begin{proposition}
If the source $X$ is standard Normal and $D(x,y) = (x - y)^2$,
then for $0 <  \Delta  \le 2$, $(R,R_c,\Delta)$ is
achievable iff 
\begin{equation}
R \ge \frac{1}{2} \log_2 \frac{1}{1 - \rho^2},
\end{equation}
where $\rho$ is the unique solution in $[0,1)$ to
\begin{equation}
\label{eq:Gauss:rho}
    1 - \frac{\Delta}{2} = \rho \sqrt{1 - 2^{-2 R_c} (1-\rho^2)}.
\end{equation}
\end{proposition}

\begin{figure}
    \begin{center}
    \scalebox{.5}{\includegraphics{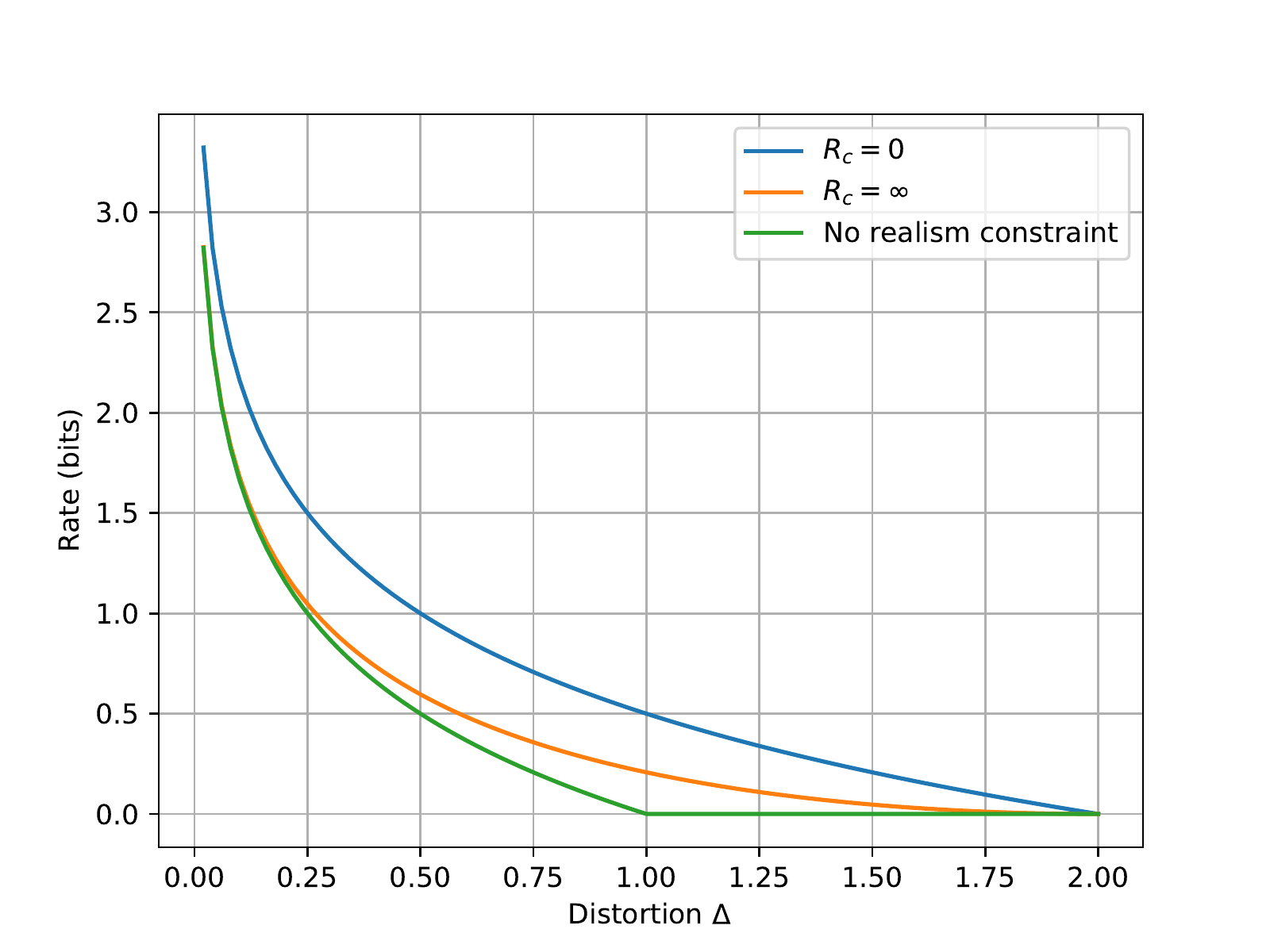}}
    \end{center}
    \caption{Rate-distortion tradeoffs for a Gaussian source
    with mean-squared error distortion for $R_c = 0$,
    $R_c \rightarrow \infty$, and the classical
    rate-distortion function without a realism
    constraint.}
    \label{fig:gauss}
\end{figure}

Taking $R_c \rightarrow \infty$ gives 
    $R \ge \frac{1}{2} \log_2 \frac{1}{\Delta(1 - \Delta/4)}$,
which was previously derived by 
Li \emph{et al.}~\cite[Prop.~2]{Li:Preserving:Long}.
On the other hand, taking $R_c = 0$ gives 
    $R \ge \frac{1}{2} \log_2 \frac{1}{\Delta/2}$,
which is consistent with the finding 
of Theis and Agustsson \cite{Theis:Advantages} and Blau
and Michael~\cite{Blau:Tradeoff} for general sources
that, under a mean squared
error distortion constraint and in the absence of
common randomness, imposing perfect realism incurs
a 3 dB penalty compared to the case without a realism 
constraint. Fig.~\ref{fig:gauss} illustrates the rate-distortion
tradeoff for $R_c = 0$, $R_c \rightarrow \infty$, 
and the classical case with no realism constraint~\cite[Sec.~10.3.2]{Cover:IT2}.
Note that at small distortions, requiring perfect 
realism incurs essentially no rate penalty, assuming
infinite common randomness is available.

\begin{IEEEproof}
    Since the source and distortion are uniformly integrable,
    we can apply Theorem~\ref{thm:main}.
Given $\rho$ satisfying (\ref{eq:Gauss:rho}), choose $U$ to be 
standard Normal so that $X \markov U \markov Y$ are jointly 
Gaussian and
\begin{align}
    E[XU] & = \rho \\
    E[UY] & = \sqrt{1 - 2^{-2R_c}(1-\rho^2)} =: \tilde{\rho}
\end{align}
with $Y$ also being standard Normal. Then we have
\begin{align}
    E[(X-Y)^2] & = E[(X - \rho U + \rho U - \tilde{\rho} U
             + \tilde{\rho} U - Y)^2] \\
    & = E[(X - \rho U)^2] + \phantom{ } \\
    & \phantom{= E[(}   E[(\rho U - \tilde{\rho} U)^2]
           + E[(\tilde{\rho} U - Y)^2] \\
    & = 1 - \rho^2 + (\rho - \tilde{\rho})^2 + 1 - \tilde{\rho}^2 \\
    & = 2 - 2 \rho \tilde{\rho} \\
    & = 2 - 2 \rho \sqrt{1 - 2^{-2 R_c} (1-\rho^2)} \\
    & = \Delta 
\end{align}
and
\begin{align}
    I(X;U) & = \frac{1}{2} \log_2 \frac{1}{1 - \rho^2} \\
    I(Y;U) & = \frac{1}{2} \log_2 \frac{1}{1 - \tilde{\rho}^2} \\
    & = \frac{1}{2} \log_2 \frac{1}{2^{-2 R_c} (1 -\rho^2)} \\
    & = R_c + \frac{1}{2} \log_2 \frac{1}{1 - \rho^2}.
\end{align}
    Thus the condition $R + R_c \ge I(Y;U)$ is equivalent
    to $R \ge I(X;U)$. Achievability then follows from 
    Theorem~\ref{thm:main}.

    To show the reverse direction, suppose $(R,R_c,\Delta)$
    is achievable. Then for all $\epsilon > 0$, 
    $(R+ \epsilon, R_c + \epsilon, \Delta + \epsilon)
      \in \mathcal{RD}$ and there exists $(U,Y)$ 
      satisfying
      \begin{align}
          Y & \stackrel{d}{=} X \\
          X & \markov U \markov Y \\
          R + \epsilon & > I(X;U) \\
          R_c + R + 2\epsilon & > I(Y;U) \\
          \Delta + \epsilon & > E[(X-Y)^2].
      \end{align}
    Now define
    \begin{align}
        \rho & = \sqrt{E[E[X|U]^2]} \\
        \tilde{\rho} & = \sqrt{E[E[Y|U]^2]}.
    \end{align}
    Then we have
    \begin{align}
        R + \epsilon & \ge I(X;U) \\
        & = h(X) - h(X|U) \\
        & \ge h(X) - h(X - E[X|U]) \\
        & \ge \frac{1}{2} \log_2 (2 \pi e)
           - \frac{1}{2} \log_2 (2 \pi e E[(X - E[X|U])^2]) \\
           & = \frac{1}{2} \log_2 \frac{1}{E[(X - E[X|U])^2]} \\
           & = \frac{1}{2} \log_2 \frac{1}{1 - E[E[X|U]^2]} \\
           & = \frac{1}{2} \log_2 \frac{1}{1 - \rho^2},
    \end{align}
    where we have used the entropy-maximizing property
    of the Gaussian distribution.
    Similarly,
    \begin{align}
          R + R_c + 2\epsilon \ge I(Y;U)
          \ge \frac{1}{2} \log_2 \frac{1}{1 - \tilde{\rho}^2},
    \end{align}
    which reduces to
    \begin{equation}
        R + 2 \epsilon \ge \frac{1}{2} \log_2 \frac{2^{-2 R_c}}{1 - \tilde{\rho}^2}.
    \end{equation}
    Turning to the distortion constraint,
    \begin{align}
        \Delta + \epsilon & \ge E[(X - Y)^2] \\
        & = E[(X - E[X|U])^2] + E[(E[X|U] - E[Y|U])^2] \\
        & \phantom{=E[(}     + E[(E[Y|U] - Y)^2] \\
        & \ge (1 - \rho^2) + (1- \tilde{\rho}^2) + \phantom{ } \\
        \nonumber
        & \phantom{\ge (1- }
              \left(\sqrt{E[E[X|U]^2]} -
              \sqrt{E[E[Y|U]^2]}\right)^2 \\
        & = (1 - \rho^2) + (1- \tilde{\rho}^2)
                 + (\rho - \tilde{\rho})^2 \\
        & = 2 - 2 \rho \tilde{\rho},
    \end{align}
    where the second inequality follows from Cauchy-Schwarz. Thus we have
    \begin{align}
        R + 2\epsilon & \ge \inf_{\rho,\tilde{\rho} \in [0,1) }
           \max \left(\frac{1}{2} \log_2 \frac{1}{1 - \rho^2},
           \frac{1}{2} \log_2 \frac{2^{-2 R_c}}{1 - \tilde{\rho}^2} \right) \\
          \text{s.t.} \ \ & \Delta + \epsilon \ge 2 - 2 \rho \tilde{\rho}.
    \end{align}
    We can assume $\Delta + \epsilon < 2$, since the $\Delta = 2$ case
    is trivial. Then at optimality, we must have
    \begin{equation}
        \frac{1}{2} \log_2 \frac{1}{1 - \rho^2}
           = \frac{1}{2} \log_2 \frac{2^{-2 R_c}}{1 - \tilde{\rho}^2},
    \end{equation}
     i.e.,
    \begin{equation}
           \tilde{\rho} = \sqrt{1 - 2^{-2 R_c}(1-\rho^2)}.
    \end{equation}
    The conclusion then follows by taking $\epsilon \rightarrow 0$.
\end{IEEEproof}

\section*{Acknowledgment}

The author wishes to thank Johannes Ball\'{e}
for introducing him to the subject
of distribution-preserving compression. 
Subsequent discussions with Johannes Ball\'{e} 
and Lucas Theis led to the results in the
paper and are gratefully acknowledged. This 
research was supported by the
US National Science Foundation under grants
 CCF-2008266 and CCF-1934985, by the US Army
 Research Office under grant W911NF-18-1-0426,
and by a gift from Google.


\end{document}